\documentclass[final,times,twocolumn]{elsarticle}

\usepackage{amssymb}
\usepackage{amsmath}
\usepackage{setspace}

\begin{document}

\begin{frontmatter}



\title{\bf The potential impact of turbulent velocity fluctuations on drizzle formation in Cumulus clouds in an idealized 2D setup\\}


\author[ox]{M. Andrejczuk\fnref{now}}
\author[le]{A. Gadian}
\author[le]{A. Blyth}

\address[ox]{Atmospheric, Oceanic and Planetary Physics,  
University of Oxford, Oxford, United Kingdom}
\address[le]{School of Earth and Environment, University of Leeds, Leeds, United Kingdom}
\fntext[now]{Now at the: 
Met Office,
FitzRoy Road,
Exeter,
EX1 3PB,
United Kingdom,
miroslaw.andrejczuk@metoffice.gov.uk
}
\begin{abstract}
This article discusses a potential impact of turbulent velocity fluctuations of the air
 on a drizzle formation in Cumulus clouds.
Two different representations of turbulent velocity fluctuations for a microphysics formulated in a Lagrangian 
framework are discussed - random walk model and the interpolation,
and its effect on microphysical properties of the cloud investigated.  

Turbulent velocity fluctuations significantly enhances velocity differences between colliding droplets, 
especially those having small sizes. As a result drizzle forms faster in simulations including a 
representation of turbulence.
Both representations of turbulent velocity fluctuations, random walk and interpolation, have similar
effect on droplet spectrum evolution, but interpolation of the velocity does account for a 
possible anisotropy in the air velocity. 

All discussed simulations show relatively large standard deviation ($\sim$1${\mu}m$) of the cloud 
droplet distribution from the onset of cloud formation is observed. Because coalesence processes 
aerosol inside cloud droplets, detail information about aerosol is available. Results from numerical 
simulations show that changes in aerosol spectrum due to aerosol processing during droplet coalescence 
are relatively small during $\sim$20 min. of the cloud evolution simulated with numerical model.

Drizzle forms initially near the cloud edge, either near the cloud top, where the mass of 
water is the largest, or near the entrainment eddies.
\end{abstract}


\end{frontmatter}


\section{Introduction}
One of the unresolved problems in cloud physics is drizzle formation in 
a warm (ice free) clouds. Early modelling studies using Lagrangian parcel
models \citeauthor{warner}~(\citeyear{warner}), \citeauthor{bartlett}~(\citeyear{bartlett}) 
demonstrated that condensational 
growth leads to a very narrow droplet spectrum, contrary to the observations, 
where the droplet spectrum inside the cloud was relatively broad 
e.g. \citeauthor{Warner1969}~(\citeyear{Warner1969}). 
A narrow droplet spectrum makes collision between droplets inefficient and as a 
result it takes a long time to switch from a condensational to a coagulational 
droplet growths. Observations show, that development of precipitation in the 
clouds may be rapid e.g. \citeauthor{goke}~(\citeyear{goke}) which couldn't be explained by the modelling results. 
Several mechanisms to explain this discrepancy between numerical model and 
observations were proposed and were reviewed by \citeauthor{beard}~(\citeyear{beard}). 

Undoubtedly representation of the cloud formation process by a parcel model 
is an approximation. Nevertheless attempts were made over the time to 
explain the width of the cloud droplet spectrum in clouds using this 
approach. \citeauthor{Hudson}~(\citeyear{Hudson}) showed that a much broader droplet spectrum can be 
obtained when instead of one vertical velocity, a velocity distribution is 
used.  Another
approach was utilized by \citeauthor{Bewley}~(\citeyear{Bewley}) and \citeauthor{lasher}~(\citeyear{lasher}), who used
a velocity field from a bulk model to derive trajectories for the parcel models. 
For both these approaches droplet spectra broader than predicted by a single 
parcel were reported, however, in these approaches the microphysics is separated 
from dynamics and thermodynamics of the Eulerian model.   

The use of the bin models \citeauthor{grabowski1989}~(\citeyear{grabowski1989}), \citeauthor{kogan1991}~(\citeyear{kogan1991}),
 \citeauthor{feingold94}~(\citeyear{feingold94}),
\citeauthor{ackerman1}~(\citeyear{ackerman1}), 
\citeauthor{khain2004}~(\citeyear{khain2004}), \citeauthor{flossmann}~(\citeyear{flossmann}), 
\citeauthor{ovchinnikov2010}~(\citeyear{ovchinnikov2010}), \citeauthor{blyth2012}~(\citeyear{blyth2012}), \citeauthor{andrzej}~(\citeyear{andrzej}),
where droplet spectrum is represented as a continuous function, made the problem 
of the transition from condensational to coagulational disappear. Even very high 
resolution (in bin sizes) parcel models with bin microphysics  are capable of producing
precipitation \citeauthor{cooper1997}~(\citeyear{cooper1997}), \citeauthor{pinsky2002}~(\citeyear{pinsky2002}) independent of whether turbulent 
enhancement is or is not taken into account.
It's not clear, however, whether the rain forms in these models due to physical processes 
or numerical errors associated 
with the numerical solution of the condensational growth and collision or the physics 
itself, since detailed comparisons between Eulerian (bin) and Lagrangian models were never to 
our knowledge reported. 

In the recent years a new approach to microphysics formulated in Lagrangian framework
was proposed for both warm rain \citeauthor{andrejczuk2008}~(\citeyear{andrejczuk2008}, \citeyear{andrejczuk2010}),
\citeauthor{shima2009}~(\citeyear{shima2009}), \citeauthor{niemcy}~(\citeyear{niemcy}) and ice clouds \citeauthor{solch}~(\citeyear{solch}) .
In this approach dynamics and thermodynamics 
are represented in the traditional Eulerian framework, whilst microphysics is represented 
in a Lagrangian framework with two way interactions between Eulerian and Lagrangian parts. 
Lagrangian microphysics tracks Lagrangian parcels (sometimes referred as super-droplets \citeauthor{shima2009}~(\citeyear{shima2009})), 
each representing a number of real 
aerosol, having the same chemical and physical properties. Depending on the conditions 
determined by the Eulerian model, water can condense on or evaporate from the aerosol surface. 
Resulting forces, together with a drag force 
are return to the Eulerian model. 
Transport of physical properties by Lagrangian parcels overcomes many problems 
present in Eulerian models. The Lagrangian representation of microphysics is diffusion 
free; each parcel can be treated individually, which makes representation of the sub-grid 
scale variability easier; the edge of the cloud is resolved without the need of the use of a 
special techniques (for instance VOF discussed in \citeauthor{reisner2}~(\citeyear{reisner2}), \citeauthor{reisner1}~(\citeyear{reisner1})). 
At the same time 
representation of the field with a
limited number of parcels may lead to random fluctuations in derived fields (for instance 
concentration of droplets and aerosol cloud water) \citeauthor{kenzelbach}~(\citeyear{kenzelbach}), \citeauthor{Salamon2006}~(\citeyear{Salamon2006});
which in an extreme cases may lead to parcel free computational grids. 

This article focuses on the representation of the turbulence in a numerical model with 
a Lagrangian representation of microphysics and its effect on drizzle formation. The effect 
of air turbulence on drizzle formation is a complex and not fully understood problem in 
cloud physics. Turbulence can affect directly relative velocities of the 
colliding droplets, leading to a coalescence of droplets which would not collide in 
a laminar flow. It can also indirectly influence drizzle formation, by modification 
of the environment in which droplets grow/evaporate. 
A broad review of the effect of turbulence on clouds and its importance
were discussed by: \citeauthor{srivastava}~(\citeyear{srivastava}), \citeauthor{jonas}~(\citeyear{jonas}),
 \citeauthor{szymon}~(\citeyear{szymon}), 
\citeauthor{Devenish}~(\citeyear{Devenish}) and \citeauthor{grabowski2013}~(\citeyear{grabowski2013}).

The effect of air turbulent velocity fluctuations on 
droplet motion in bin models may be represented as an enhancement of collision efficiencies 
derived from a high resolution turbulence simulations \citeauthor{franklin}~(\citeyear{franklin}), \citeauthor{pinsky2006}~(\citeyear{pinsky2006}),
\citeauthor{wang}~(\citeyear{wang}). This method has been used recently by 
\citeauthor{pinsky2002}~(\citeyear{pinsky2002}), \citeauthor{andrzej}~(\citeyear{andrzej}).  
Also, a recently reported model with Lagrangian microphysics \citeauthor{niemcy}~(\citeyear{niemcy}) uses this approach. 
Because in the Lagrangian microphysics the
parcel velocity is a predicted variable, it can be used with the
parameterization of the sub-grid scale velocity fluctuations to determine the relative velocity of the colliding parcels.

This article reports an application of a two possible representations (parameterizations) of turbulence for a model with 
microphysics formulated in a Lagrangian framework, one - random walk model, and the other
interpolation of the velocity to parcel location and investigates its effect 
on a droplet spectrum. The results from these models are compared with the results 
from the model where parcel terminal velocity is used when calculating probability of
droplet collisions. 
When for a parcel predicted by the model velocity is used to calculate probability
of droplet collisions it is assumed that droplet turbulent transport and 
interaction occur in the same scales, because the same velocity is used for both.
Assumption that colliding droplets have terminal velocity corresponds to the case, 
where turbulent transport and interaction (collision-coalescence) between droplets 
happen in a different scales.
This gives a lower and an upper limit of the effect of the turbulence on droplet collisions, because 
in former case it is consistent with the turbulent transport model and in latter is neglected.

The next section describes numerical model and representation 
of turbulence.  Results are discussed in section 3, and 
conclusion are in the last section, 4. 

\section{Numerical Model}
\subsection{Eulerian Model}
In this article the EULAG model is used as a driving model for Lagrangiam microphysics. 
Eulag is an Eulerian/Semi-Lagrangian solver (\citeauthor{Smolarkiewicz1997}~(\citeyear{Smolarkiewicz1997}), 
\citeauthor{Prusa2008}~(\citeyear{Prusa2008})), with the Eulerian version used in the simulations reported in this article.
Model equations in an anelastic 
approximation can be written as: 
\begin{equation}
\frac{\partial\rho\psi}{\partial t}+\nabla\rho v\psi=\nabla\rho K_{m}\nabla\psi+\rho F_{\psi}+\rho F_{p}
\label{eq1}
\end{equation}
where $\rho$ is density, $K_m$ diffusion coefficient, $\psi$ - 
any dependent variable (u, w, $\theta$, $q_v$) with its associated 
Eulerian forcing - $F_{\psi}$, $F_p$ describes forces from a Lagrangian 
model and are defined below. $K_m$ is derived from a prognostic 
TKE equation \citeauthor{dirdoff2}~(\citeyear{dirdoff2}), see details about implementation in \citeauthor{margolin}~(\citeyear{margolin}).
\subsection{Lagrangian Microphysics}
Lagrangian microphysics tracks Lagrangian parcels, each representing a group of 
real aerosol particles. Each parcel can be characterized by the same velocity and size 
(both dry aerosol radius and radius of the droplet if water condenses on the parcel), 
and occupy approximately the same place in a physical space. 
 
Lagrangian model equations describing evolution in time of the droplet radius ($r_i$), 
velocity ($v_i$) and location ($x_i$) for a parcel $p$ can be written as:
\begin{equation}
\frac{dr_p}{dt} =\frac{G}{r_p} (S-S_{eq}) 
\label{r_grow}
\end{equation}
\begin{equation}
\frac {{du_p}_i} {dt} = \frac{1}{{\tau}_i} (u^d+u'-{u_p}_i) + {\delta}_{i2}g 
\label{u_eq}
\end{equation}
\begin{equation}
\frac{dx_p}{dt}=v_p 
\label{x_eq}
\end{equation}
where  $r_p$ -droplet radius, $S$ - supersaturation at parcel location, $S_{eq}$ - equilibrium 
supersaturation for given aerosol size and temperature 
(see \citeauthor{andrejczuk2008}~(\citeyear{andrejczuk2008}) for detail description of other symbols). 
The $u^d$ -  deterministic air velocity 
and 
$u^{'}$ - turbulent component determined from the sub-grid scale model; $g$ is gravity. 
Equation \ref{r_grow} is solved using VODE solver (\citeauthor{Brown}~(\citeyear{Brown})), equation \ref{u_eq} is solved 
using the backward Euler method, and equation \ref{x_eq} is solved using the forward Euler method.

One of the established ways to represent turbulent component of the flow velocity is by 
a random process (random walk) having
normal distribution and standard deviation $\sqrt{2K_{L}/\Delta{t}}$, which is derived 
in the Appendix. Note that to determine $K_{L}$ different mixing length than to 
determine $K_m$ in eq. \ref{eq1} was used. 
Although the mixing length of the order of model grid size is 
typically used in sub-grid scale model, in the Lagrangian model because the location of each 
parcel within the grid is known, the mixing length based on $TKE$ and $dt$ can be derived. 
This, simplified, description of the turbulence treats turbulence as a random walk 
process and represents the effect of air turbulent velocity on Lagrangian parcels velocity through 
the variability in the Eulerian model velocity field within a computational grid. 
A broad review of theory and application
of Lagrangian transport models can be found in \citeauthor{wilson}~(\citeyear{wilson}).
No attempt has been made to represent eddies present in the turbulent 
flow, but not resolved by the model;  it is assumed that at each time-step 
turbulent fluctuations of the air velocity are independent. It is achieved by generating a new
random number for each Lagrangian parcel on each time-step. This method follows 
an approach proposed by \citeauthor{almeida1976}~(\citeyear{almeida1976}). 

The other method of representing variability of the velocity
within the computational grid is to use interpolation; and instead of using
the same value of the velocity for all parcels within a given computational
grid the velocity can be interpolated from neighboring grids to each parcel location.  

These two approaches, later referenced as a turbulence,  
provide a way to describe the variability of the air turbulent velocity inside 
computational grids in the spirit discussed by \citeauthor{srivastava}~(\citeyear{srivastava}) for a microscopic 
supersaturation. 

Details of representation of the coalescence process in Lagrangian microphysics 
is described in  detail in
\citeauthor{andrejczuk2010}~(\citeyear{andrejczuk2010}) and \citeauthor{andrejczuk2012}~(\citeyear{andrejczuk2012}). 
Collisions of all Lagrangian parcels within the collision grid (this does not have to be 
the same as Eulerian grid, in the simulations reported in this article collision grid 
is a quarter of a computational grid), with sizes larger than 3 ${\mu}$m and having water 
on the surface are considered, and \citeauthor{long1974}~(\citeyear{long1974}) analytical expression for gravitational collision efficiency is used. Each 
collision event, based on the size of colliding droplets and aerosol size inside the 
droplets is assigned to one of the pre-defined microphysical grids spanning both aerosol 
sizes and droplet sizes. For each microphysical grid the mass of the aerosol, the mass of the 
water and the number of new droplets is calculated. Based on this information new parcels 
are created for each microphysical grid for which the number of physical droplets is larger 
than specified threshold level - 156.25 in simulations discussed in this article (which 
corresponds to resolving 1 droplet/m$^3$).  Newly created parcels 
for collision grids at the edge of the 
cloud are placed next to randomly chosen parcels within this grid. For grids within the 
cloud they are randomly placed within the Eulerian computational grid. Additionally the algorithm 
assures also that existing parcels represent a larger than threshold level number of 
physical particles. Should the collision lead to a smaller number, the probability of collision, 
with all other parcels for this particular one is reduced. This representation of 
coalescence process allows to process not only droplet sizes, but also aerosol sizes during 
droplet coalescence. It is assumed that newly created Lagrangian parcels move with terminal velocity. 

\subsection{Models coupling}
Drag force and tendencies for a temperature and water vapour mixing ratio 
are calculated in Lagrangian model and used in an Eulerian part:
\begin{equation}
Fp_{\bf{u}}=\frac{1}{\Delta V}\sum_{p~\in~grid_{(i,j,k)}} \frac{M_p}{{\tau}_p} ({{\bf{u}}_p}-u^{*}) 
\end{equation}
\begin{equation}
Fp_{qv}=\frac{4 \pi {\rho}_p}{3\Delta V}\sum_{p~\in~grid{(i,j,k)}} M_p(r_p^3-r_{0_p}^3)
\end{equation}
\begin{equation}
Fp_{\theta}=  - \frac{{\theta}_e~L}{T_e~c_p}~Fp_{qv}, 
\end{equation}
where indexes $i$, $j$, $k$ index numerical model computational grids, $M_i$ is the number 
of real particles parcel $i$ represents. A $T_e$ and ${\theta}_e$ are 
temperature and potential temperature profiles. 
Treatment of these forces is similar to the treatment of the sub-grid scale tendencies as
discussed by \citeauthor{margolin}~(\citeyear{margolin}).
A finite-difference approximation to eq. \ref{eq1} can be written as (\citeauthor{margolin}~(\citeyear{margolin})):
\begin{equation}
{\psi}^{n+1}_i=ADV(\tilde{\psi}_i)+0.5{\Delta}tF^{n+1}_i,
\end{equation}
where $\tilde{\psi} = {\psi}^n_i+0.5{\Delta}tF^n_i+{\Delta}tD_{\psi}+{\Delta}tFp_{\psi}$,
$F^{n+1}_i$ are forces associated with pressure gradient, absorbers, and buoyancy
(\citeauthor{Smolarkiewicz1997}~(\citeyear{Smolarkiewicz1997}), \citeauthor{margolin}~(\citeyear{margolin})). $ADV$ - denotes advection, which is calculated 
using MPDATA scheme (\citeauthor{Smolarkiewicz1983}~(\citeyear{Smolarkiewicz1983}), \citeauthor{Smolarkiewicz1990}~(\citeyear{Smolarkiewicz1990}).
\subsection{Model Setup}
In this article a 2D idealized setup is used to investigate the effect of air turbulent velocity on drizzle formation. 
Initially, an atmosphere with a $1/\theta d \theta/ dz$ = 1.3 $m^{-1}$ was specified. Below 2km a relative 
humidity was defined as 85\%, dropping to 75\% above this level. The model domain covers 3.2 km $\times$ 5.0 km, 
resolved with 25m resolution in each direction. Three-modal, log-normal aerosol distribution, 
corresponding to continental air, have been specified for the whole domain using values from  
\citeauthor{withby}~(\citeyear{withby}): 
$r_n$ =0.008 ${\mu}$m, ${\sigma}_n$=1.6, $N_n$=1000, 
$r_a$ =0.034 ${\mu}$m, ${\sigma}_a$=2.1, $N_a$= 800, 
$r_c$ =0.46 ${\mu}$m, ${\sigma}_c$=2.2, $N_c$=0.72. 

\begin{figure*}[tbh]
\centerline{\includegraphics[width=6.0in,clip=true]{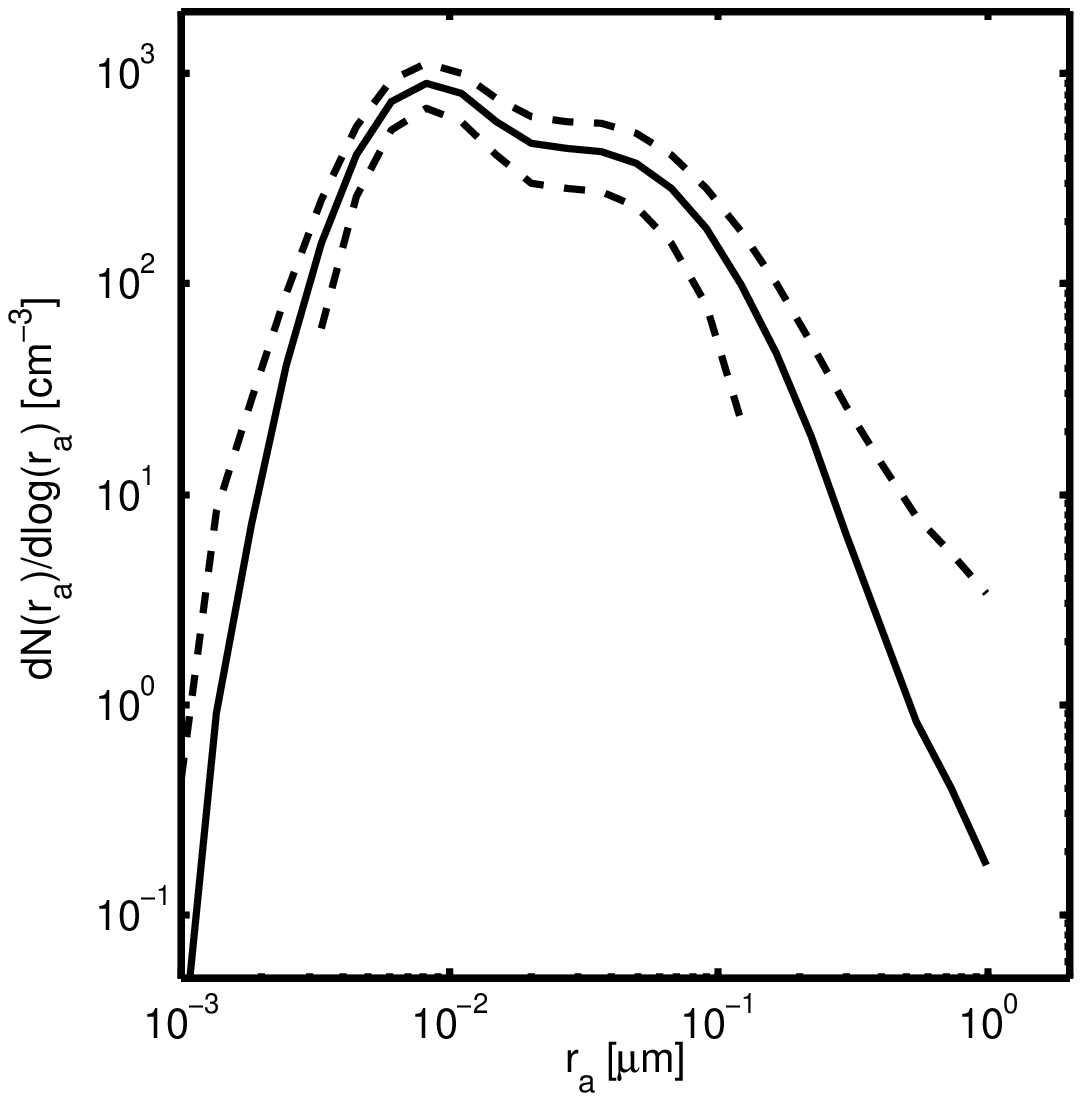}}
\caption{Averaged over the whole domain initial aerosol distribution. Dashed lines show
standard deviation of the number for each aerosol bin.} 
\label{fig1}
\end{figure*}

An initial aerosol distribution in each computational grid
was represented by 100 parcels.
The aerosol spectrum used in the simulations is shown in figure \ref{fig1}. 
Aside from a spectrum averaged 
over the whole domain, the standard deviation of the values for each bin 
is also plotted. Variability in each bin
is a result of a random sampling of the initial distribution with finite 
number of parcels. Although on average 
the spectrum agrees with an analytical distribution, sampling with a limited number 
of parcels leads to a variability
of the distribution for a different grids and as a result it affects droplet 
number and size in each model grid during 
the condensational growth. 

The model was forced by the surface temperature source prescribed as:
\begin{equation}
f_t(x,z)=A_{th}exp(-\frac{(x-x_0)^2+(z-z_0)^2}{800}),
\end{equation}
with $A_{th}$=0.15 K/s and ($x_0$, $z_0$) set to (0,400).  
Simulations have been run for 1080s, with a time-step 0.25s.   
Microphysical grid in aerosol space (in ${\mu}m$) was specified as:
$r_a(k)=10^{-3.2+k*0.2}$, with 30 bins. In radius space 28 bins were used,
and radius for a bin $k$ was specified as:
$r(k)=p^{1/3}r(k-1)$, with r(1)=1${\mu}m $ and $p$=2.

The following model setups are discussed in details in this article: 
\begin{description}
	\item {$REF$} - reference simulation. No representation of turbulence 
         in eq. \ref{u_eq} (v${'}$=0). Deterministic flow velocity u$^d$ at parcel 
	 location  
	 is determined by interpolating velocity from 4 (8 in 3D) nearest
	 Eulerian computational grids to a parcel location. The vertical component of the 
	 parcel velocity is used
	 to evaluate collision kernel.
\item {$BIN$} - As far as parcel movement is concerned 
	it is assumed that parcel velocity is equal to the air velocity 
	interpolated to a parcel location,
	that is eq. \ref{u_eq} is not solved. The terminal velocity of the parcel
	is added to the vertical velocity. 
	The parcel's terminal velocity has been determined from the expression 
        given by \citeauthor{simmel2002}~(\citeyear{simmel2002}). To simplify calculations instead of the terminal 
	velocity for a parcel size, terminal velocity for the center of the collision bin
	to which the parcel is assigned is used.
	When the collision kernel between parcels is calculated it is 
        assumed that the parcel velocity is equal only to a terminal velocity. 
        This setup mimics bin model as far as collision-coalescence process is concerned and assumes
	that turbulent transport and collision happen in different scales (are independent). 
\item {$TURB$} - for a given parcel deterministic flow velocity has 2 components
	deterministic:
        $u^{d}= u({\bf x})- \frac {1} {\rho} \frac{\partial} {\partial x_{i}} (\rho K_{l})$, where
	u({\bf x}) is Eulerian velocity for the computational 
	grid to which parcel belongs, $K_l$ - diffusion coefficient for a Lagrangian model, 
	and $\rho$ - gas density;
	and turbulent component v${'}$=$\sqrt{\frac{2K_l}{\triangle t}}$
	with the derivation of of this model presented in
        the Appendix. 
\item {$REF2$} -  Similar to $REF$, but in this case in additional to the 
         difference in vertical velocity also difference in
	 horizontal velocity is taken into account when evaluating the collision kernel.
\end{description}

\section{Results}
\subsection{Evolution in space}
\begin{figure*}[tbh]
\centerline{\includegraphics[width=6.0in,clip=true]{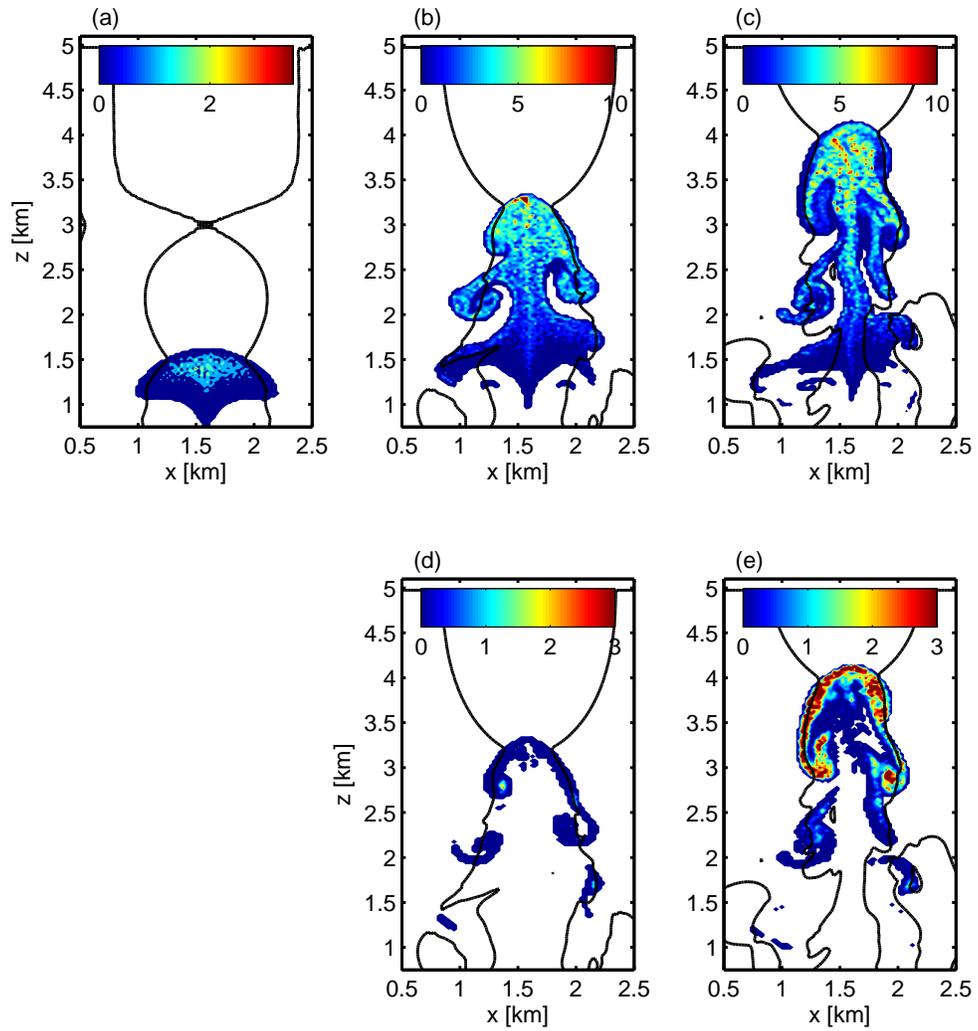}}
\caption{ Upper row - cloud water mixing ratio (q$_c$) after 8 min. (a), 15 min (b) and 17 (c) min.;
	lower row - rain water mixing ratio (q$_c$) after 15 min. (d), 17 min (e)  for a $REF$ run. 
	Solid black line shows contour for value 0 of the vertical velocity.
} 
\label{fig2}
\end{figure*}

Figure \ref{fig2}a-\ref{fig2}c, shows snapshots of a cloud water mixing ratio ($q_c$) for the model 
solution for times 8 min. (fig. \ref{fig2}a), 15 min (fig. \ref{fig2}b) and 17 min 
(fig. \ref{fig2}c) for the $REF$ run.  Cloud evolution exhibits features typically found 
in a 2D Cumulus developing in a stable stratified atmosphere and reported already by: 
\citeauthor{klassen1985}~(\citeyear{klassen1985}), \citeauthor{grabowski1991}~(\citeyear{grabowski1991}), \citeauthor{brenguier1993}~(\citeyear{brenguier1993}). 
After reaching condensation level water vapor condenses and forms the cloud; air continues moving 
upward due to the temperature excess within the thermal. During the ascent in a stably 
stratified air cloud mixes with the environmental air, forming entrainment eddies. The cloud 
water mixing ratio reaches values as large as 12 g/m$^3$ during the simulation, and at the 
end of the simulation the maximum vertical velocity is around 18 m/s. The cloud water mixing ratio 
shows relatively large variability in space, being a result of the fluctuations of the parcel 
number within the computational grid and also because of the variability in the 
randomly generated aerosol spectrum for each model grid.  
  
Because in the numerical model the coalescence process is present, coalescence of the droplets in time 
leads to a drizzle formation. 
Figure \ref{fig2}d and \ref{fig2}e show evolution in time of the q$_r$ - the rain water mixing 
ratio for a $REF$ case. A 50 ${\mu}m$ droplet radius is defined to be a border between a cloud and a 
rain droplet sizes. Large droplets reside near the edge of the cloud, and at the end of the 
simulation $q_r$ exceeds 3 g/kg. This behavior is typical for the three setups:  
$REF$, $TURB$ and $REF2$. For a $BIN$ case negligible q$_r$ forms.

The edge of the cloud is a place where the difference in the vertical velocity within the collision 
grid is the largest (either because of the gradient of velocity between the interior and exterior 
of the cloud for $REF$ or $REF2$, or because TKE, used to derive stochastic velocity perturbations is 
largest). Additionally at the edge of the cloud the droplet spectrum can be broader than in the core 
of the cloud, which also enhances coalescence between droplets. The width of the droplet spectrum 
increases near the edge of the cloud because: a) the gradient of the supersaturation is largest there
and as a result the interpolation of the thermodynamical parameters 
to a parcel location within the same computational grid different droplet 
populations may grow and evaporate at the same time depending on the distance from the edge 
of the cloud; b) vertical velocity changes sign near the cloud edge (in figure \ref{fig2} the solid 
line shows a contour of the value 0), as a result
droplets evaporate in a down-drafts near the cloud edge; and c) because largest droplets, formed 
in the center of the up-draft, after reaching cloud top start moving along cloud edge. 
\subsection{Evolution in time}
\begin{figure*}[tbh]
\centerline{\includegraphics[width=6.0in,clip=true]{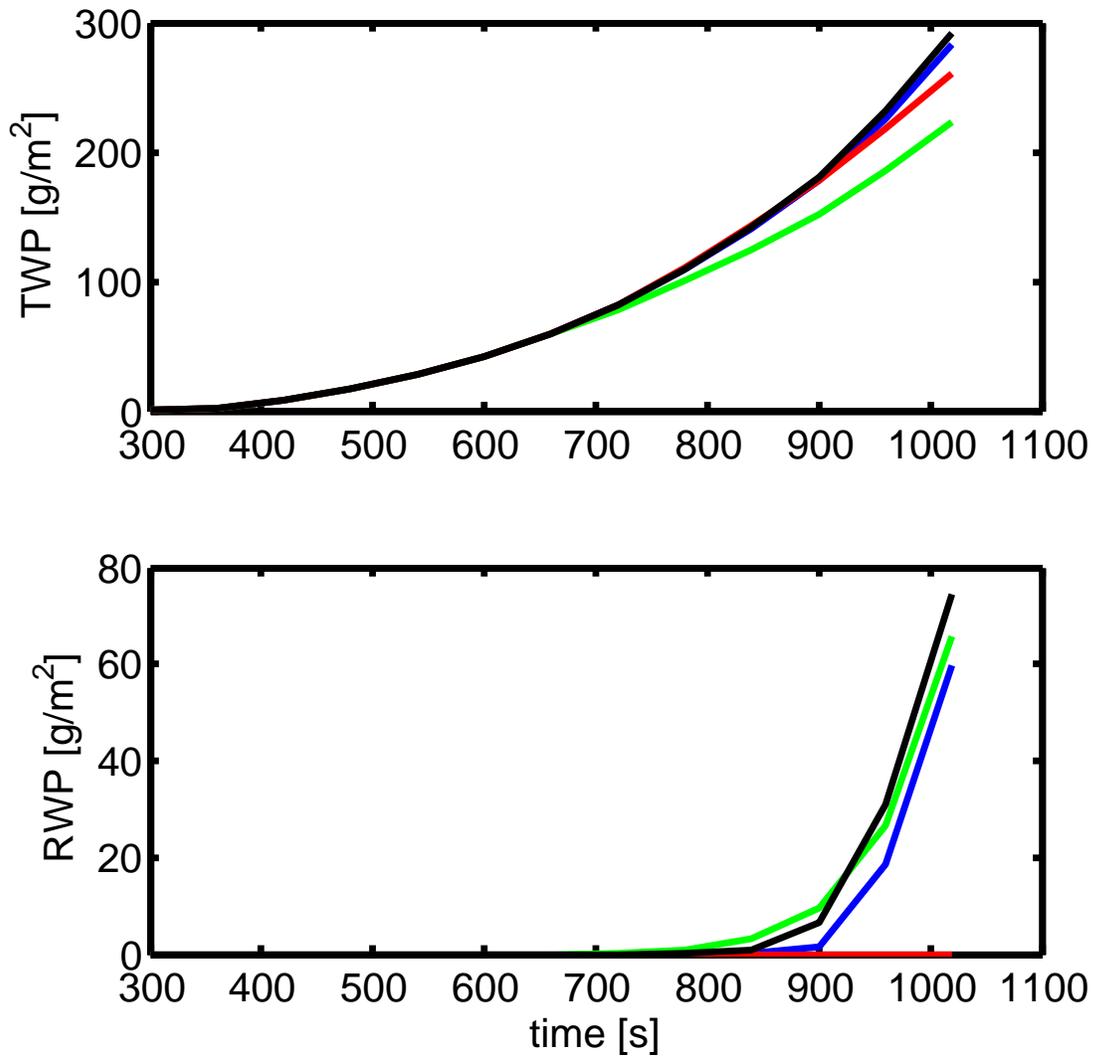}}
\caption{Evolution in time of a Total Water Path (a) and a Rain Water Path (b) for
$REF$ - blue, $BIN$ - red, $TURB$ - green, $REF2$ - black.
}
\label{fig3}
\end{figure*}

Evolution of the TWP (Total Water Path) and RWP (Rain Water Path) in time for cases discussed 
are shown in figure \ref{fig3}.  During the early stage of the cloud development  TWP is very 
similar for all cases. In time, however, differences appear as a result of the interaction of
the cloud and flow dynamics. The largest amount of water is for $REF2$ and $REF$ and smallest for 
$TURB$. The decrease in a TWP for a $TURB$ case can be associated with a random velocity of the 
air within the grid. As a result evaporation at the edge of the cloud may be larger for this 
case, because droplet trajectories can be different than for other three cases when velocity 
is interpolated to a parcel location. Much larger differences are observed for the RWP. 
The largest amount of water is for the $REF2$ case, the simulation where in coalescence 
calculation full velocity is taken into account. Compared to $REF$ simulation, $REF2$ has 
around 15\% more RWP.  For the $BIN$ simulation negligible amount of RWP has formed during 
the simulation time. But undoubtedly the coalescence process is active for this case also and 
droplets as large as 40 ${\mu}m$ have been formed in this simulation. These droplets are much 
smaller than for the three other simulations, where sizes up to 370 ${\mu}m$ are present.      
\subsection{Vertical profiles}
\begin{figure*}[tbh]
\centerline{\includegraphics[width=6.0in,clip=true]{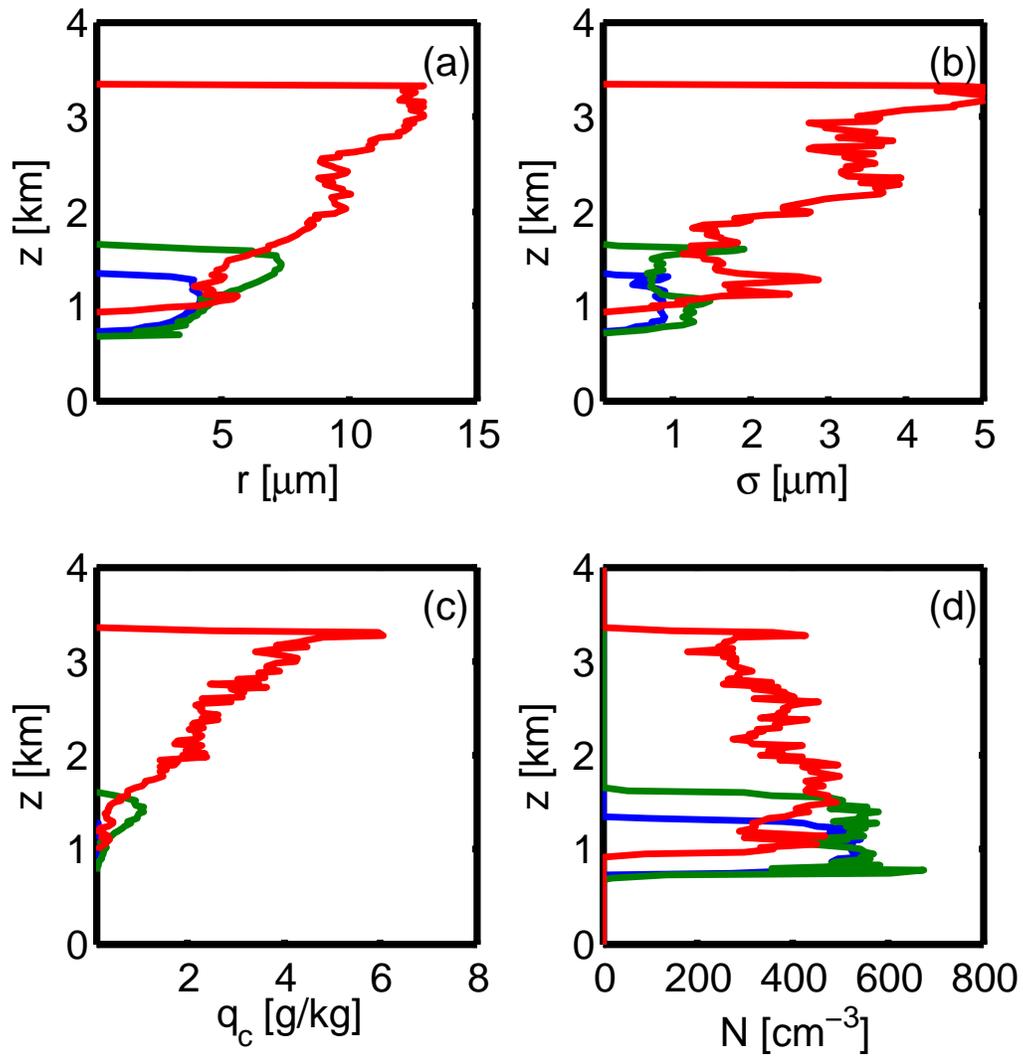}}
\caption{Vertical profiles for a $REF$ case of mean radius (a), 
standard deviation of the droplet spectrum on given level  (b),
q$_c$ (c), and cloud droplet concentration (d) for times
6 min. -  blue, 8 min. - green, 15 min. - red.
}
\label{fig4}
\end{figure*}

\begin{figure*}[tbh]
\centerline{\includegraphics[width=6.0in,clip=true]{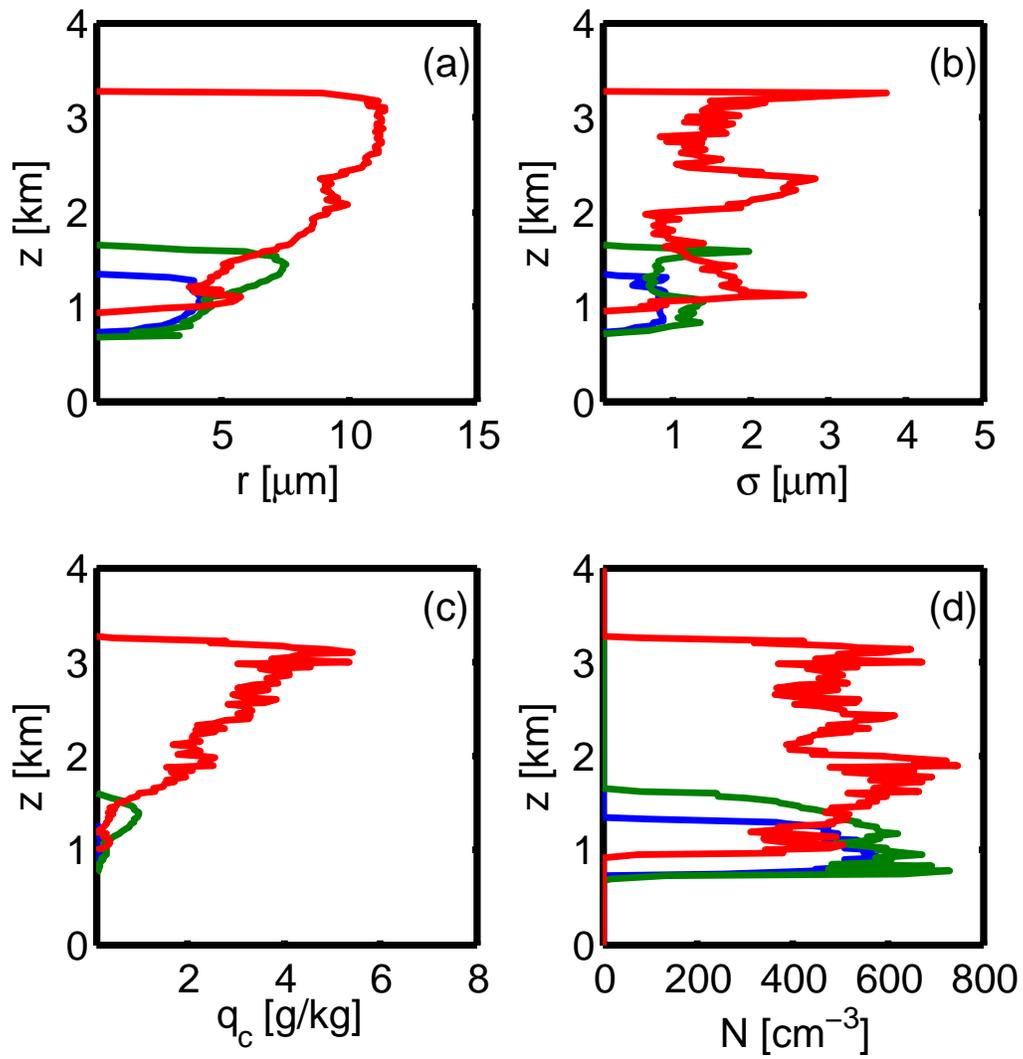}}
\caption{As figure \ref{fig4}, but for a $BIN$ case.
}
\label{fig5}
\end{figure*}

Vertical profiles of the mean radius ($r$), standard deviation of cloud droplet 
distribution ($\sigma$), cloud water mixing ratio ($qc$) and number of cloud droplets are 
shown in figures \ref{fig4} for a $REF$ case (with very similar statistics observed for a 
$REF2$ and $TURB$ simulations) and for a $BIN$ case in figure \ref{fig5} for a 3 times: 6, 
8, and 15 minutes from the beginning of the simulation.  These 3 times show cloud 
characteristics at the very early stage of the cloud development (the time when  
$q_c$ becomes larger than 0), through the initial formation of the drizzle 
(space distribution after 8 min. is shown in figures \ref{fig2}b and \ref{fig2}d), 
to the moment when significant drizzle develops (plots \ref{fig2}c and \ref{fig2}e). 
These diagnostics were calculated for each model level by taking into account only model grids, 
where $q_c$ was larger than 10$^{-3}$ g/kg. Figures \ref{fig4} and \ref{fig5} show a very similar development of the cloud 
for both cases. Initially a small amount of water condenses, but at the same time the mean radius 
is around 5 ${\mu}m$ and standard deviation around 1 ${\mu}m$. 
Around 1/3 of the total aerosol concentration
activates at 
this stage. In time, the cloud thickens and after 8 minutes the cloud base moves upward. The mean droplet 
radius increases with height and reaches 12-13 ${\mu}m$ near the cloud top after 15 minutes 
At that time the number
of cloud droplets decreases with height for a $REF$ case from  500 cm$^{-3}$ to 250 cm$^{-3}$.
For the $BIN$ simulation, the cloud droplet 
concentration does not change significantly with height and oscillates around 500 cm$^{-3}$.  
Another difference between these 2 cases is in the standard deviation of the cloud droplet 
distribution after 15 min. above 2.5 km, and for the $BIN$ case standard deviation  
is around 1.5 ${\mu}m$, except at the cloud top, where it 
reaches 4.2 ${\mu}m$. For $REF$ case it is of the order of 3.5 ${\mu}m$, with the value 
near the cloud top of $\sim$5 ${\mu}m$. 
At earlier times and below 2.5 km standard deviation is very 
similar for both cases.
\subsection{Droplet and aerosol spectra}
\begin{figure*}[tbh]
\centerline{\includegraphics[width=6.0in,clip=true]{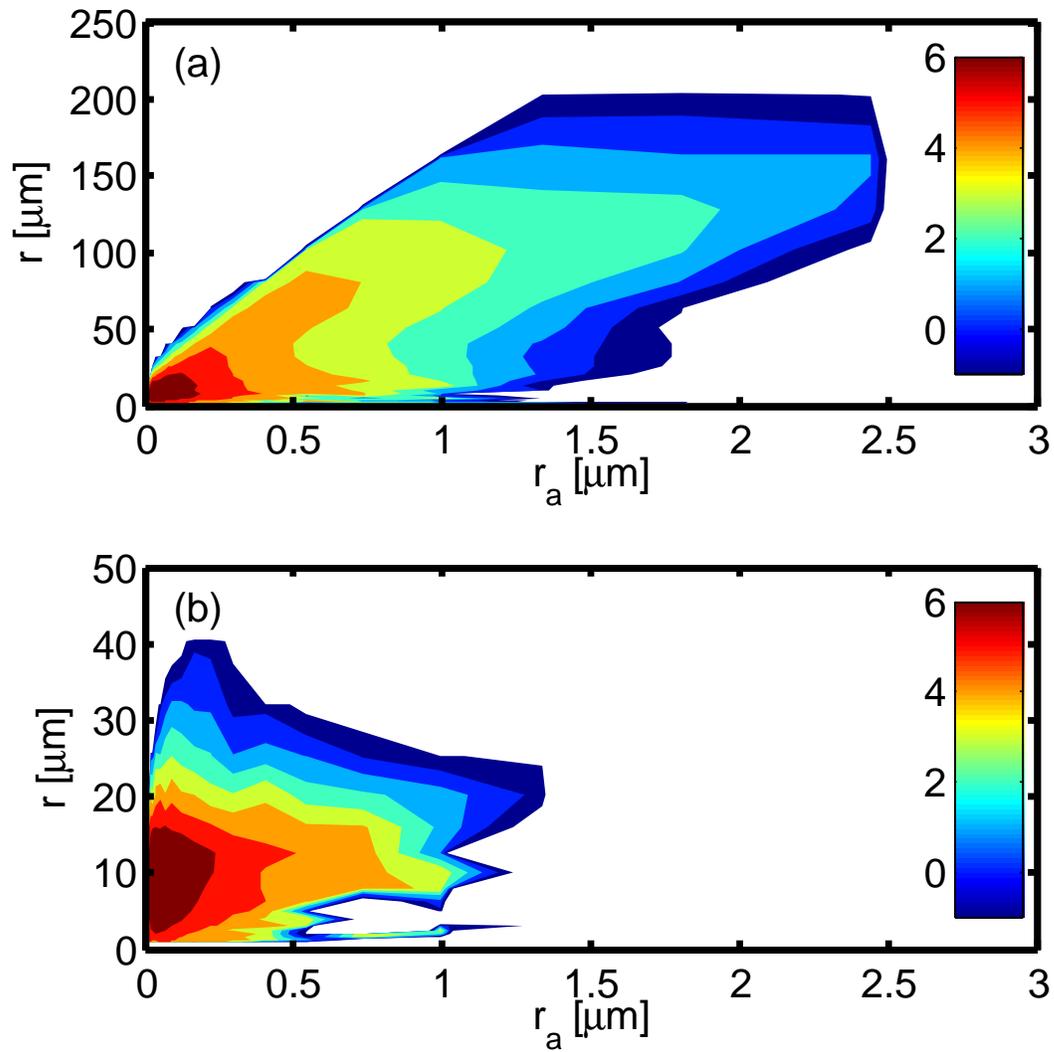}}
\caption{Relation between aerosol size and droplet size for grids with
q$_c > 10^{-3}$ - Log$_{10}$(N(r,r$_a$)) for run $REF$ (a) and $BIN$ (b).
}
\label{fig6}
\end{figure*}

Because for each parcel information about droplet size and aerosol size is available,
the relation between aerosol sizes and droplet sizes can be derived. Figure \ref{fig6}
shows this relation mapped on a Eulerian microphysical grid after 18 minutes. 
The relation between aerosol size and droplet size is complex and for a given 
aerosol size there is a broad range of droplet sizes formed on it. Because initially 
aerosol sizes were limited to 1 ${\mu}m$, sizes larger than that were created in a 
coalescence process. Much larger sizes of both aerosol and droplet are present for a $REF$
case, where the coalescence is more intensive. 

\begin{figure*}[tbh]
\centerline{\includegraphics[width=6.0in,clip=true]{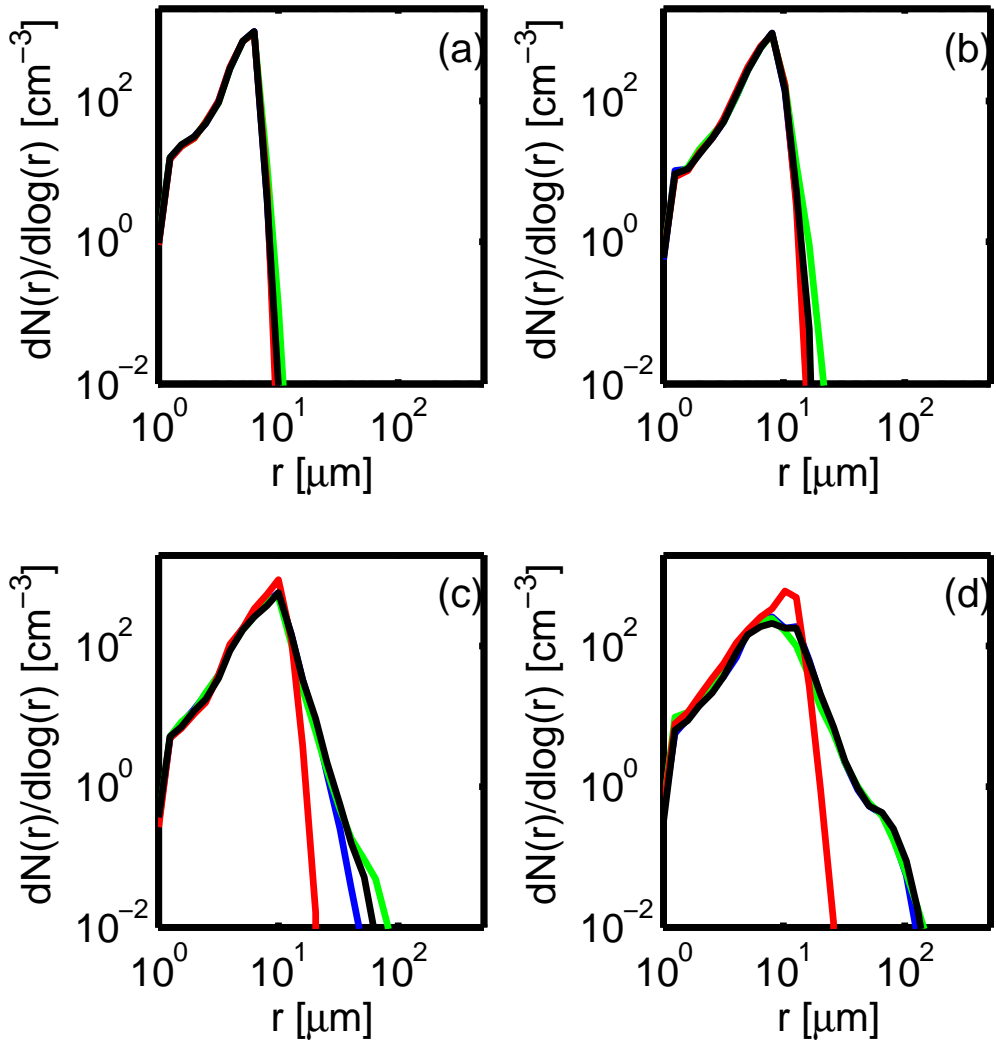}}
\caption{ Droplet spectrum after 8 min. (a), 12 min. (b), 15 min (c) and 18 min (d)
for 
a $REF$ - blue, $BIN$ - red, $TURB$ - green, $REF2$ - black.
}
\label{fig7}
\end{figure*}

From observations or spectrum resolving (bin) models typically information about either
aerosol distribution or droplet distribution is derived. This information for
microphysics in a Lagrangian framework is obtained by integrating relation
shown in figure \ref{fig6} along one of the dimensions.
Averaged over the whole cloud the droplet spectrum is shown in figure \ref{fig7}. After 
9 minutes - figure \ref{fig7}a the droplet spectrum is very similar for all cases. 
In time, similar to other cloud properties 
the differences between $BIN$ and $REF$, $TURB$ and $REF2$ emerge 
(\ref{fig7}b, \ref{fig7}c, \ref{fig7}d). 
Formation of the large droplets for the $BIN$ 
simulation is much slower than for the remaining three cases. In an early stage of 
cloud development the largest droplets form for the $TURB$
case. At the end of the simulation, the spectra
for $REF$, $TURB$ and $REF2$ cases are very similar.

\begin{figure*}[tbh]
\centerline{\includegraphics[width=6.0in,clip=true]{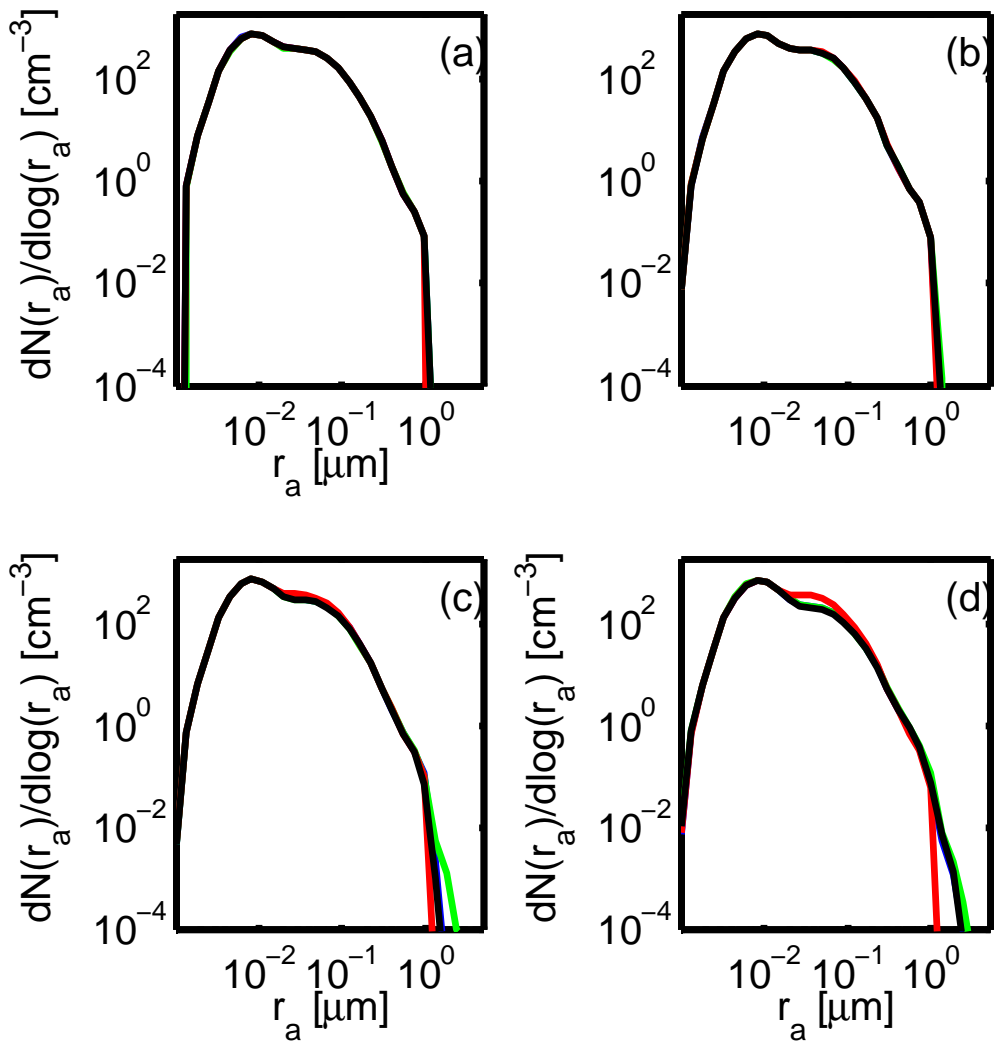}}
\caption{As figure \ref{fig7} but for aerosol spectrum.
}
\label{fig8}
\end{figure*}

Evolution of an aerosol spectra averaged over the whole cloud is presented in figure \ref{fig8}. 
Because coalescence is processing not only droplet sizes, but also aerosol sizes, 
in time the aerosol spectrum also 
changes. There are indications, especially for later times, that aerosol is processed during droplet 
coalescence, 
but the differences between $BIN$ (where aerosol processing is negligible) and other three 
simulations are small. Fastest aerosol processing is for a $TURB$ simulations, 
with a smaller, for a $REF$ and $REF2$. 
Much more time is needed for the coalescence to process aerosol and form 
large/giant aerosol than the length of an idealized simulation discussed in this article. 
There is, however, 
evidence that this aerosol processing (eg. \citeauthor{cynthia}~(\citeyear{cynthia})) may be an important source of large aerosol, because within 
the $\sim$20 minutes maximum aerosol size has doubled, 
reaching 2.5 ${\mu}m$ radius (see fig. \ref{fig6}a) 
compared to 1 ${\mu}m$ initially. The increase of the concentration of the large aerosol is at the expense
of the aerosol having sizes between 0.01 and 0.2 ${\mu}m$.
\subsection{Velocity enhancement}
\begin{figure*}[tbh]
\centerline{\includegraphics[width=6.0in,clip=true]{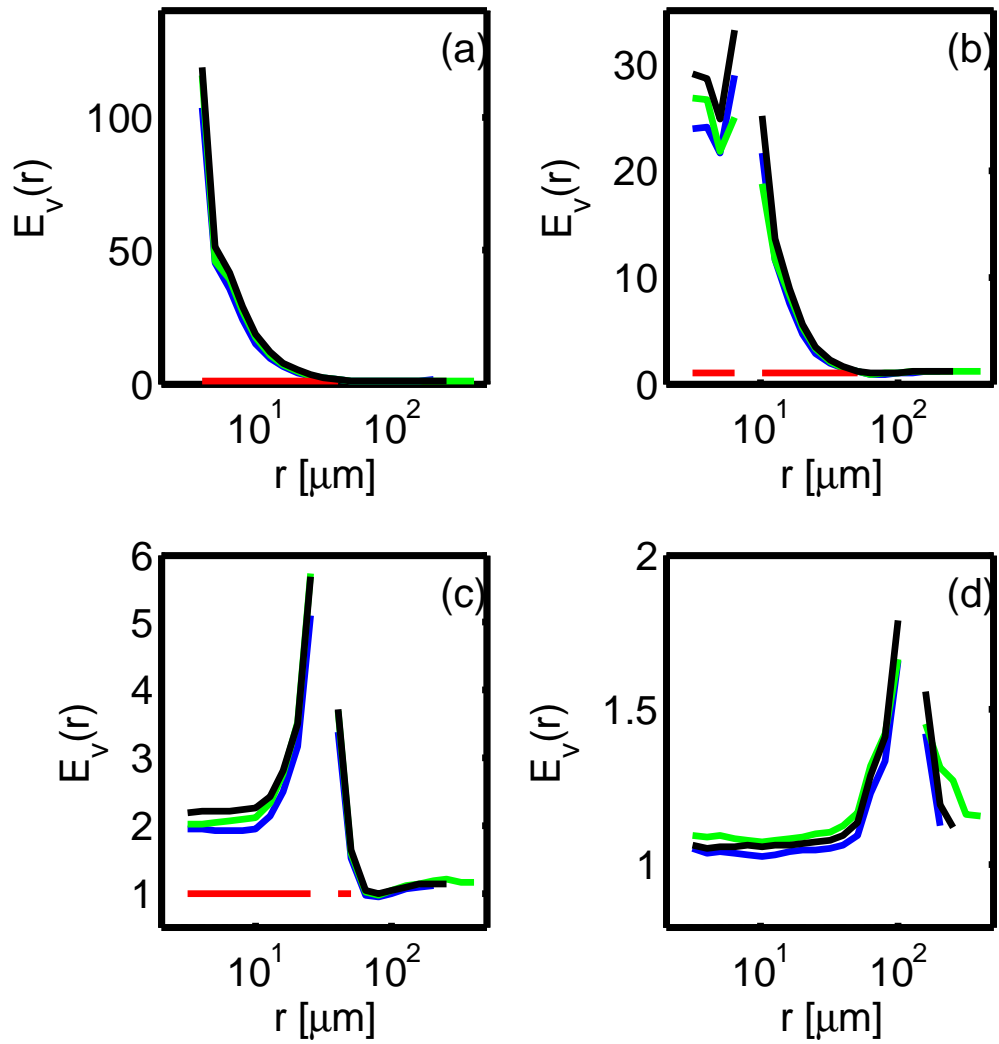}}
\caption{A velocity enhancement factor (E$_v$)for a different microphysical bins:
bin 4 (a), bin 10 (b), bin 16 (c), bin 22 (d) for a 
$REF$ - blue, $BIN$ - red, $TURB$ - green, $REF2$ - black.
}
\label{fig9}
\end{figure*}

\begin{figure*}[tbh]
\centerline{\includegraphics[width=6.0in,clip=true]{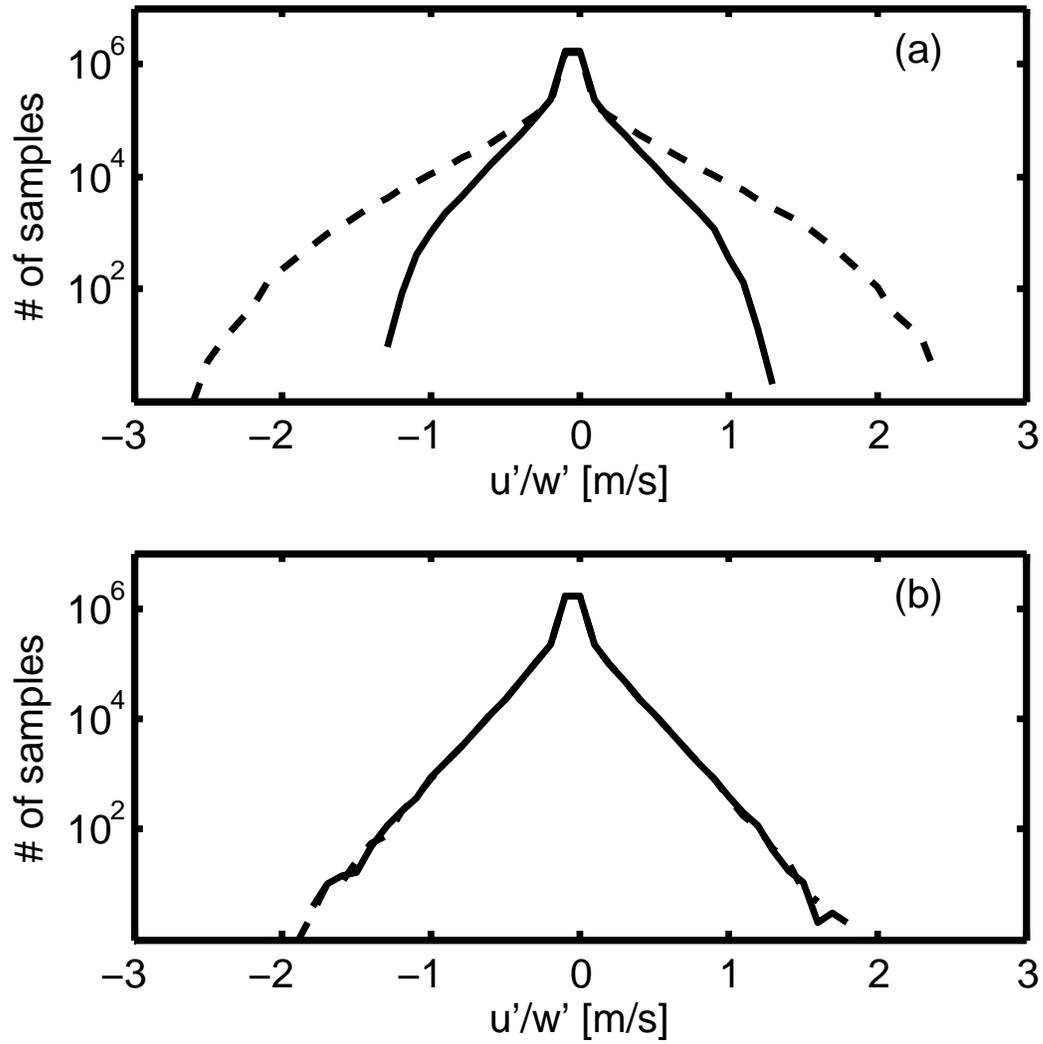}}
\caption{Distribution of the velocity fluctuations within the computational grid
for a $REF$ (a) and a $TURB$ (b) case. Solid line $u^{'}$, dashed line - $w^{'}$.
}
\label{fig10}
\end{figure*}

The collision kernel, describing the probability of the coalescence of two colliding droplets 
can be written as:
\begin{equation}
K(r_i,r_j,v_i,v_j) =\pi (r_i+r_j)^2 |v_i-v_j|E_{i,j},
\label{r_call}
\end{equation}
where r$_i$,r$_j$ is the radius of a droplet $i$/$j$ with the corresponding velocity v$_i$/v$_j$,
E$_{i,j}$ is the gravitational collision efficiency. 
With the equation of motion solved for each parcel in the model, diagnostics of
the relative velocity of colliding droplets can be determined. 
Equation
\ref{r_call} can be rewritten in the following way:
\begin{equation}
	K(r_i,r_j,v_i,v_j) =\pi (r_i+r_j)^2 \frac{|v_i-v_j|}{|w_i-w_j|} |w_i-w_j| E_{i,j},
\label{r_call1}
\end{equation}
with the w$_i$/w$_j$ being the terminal velocity for a parcel $i/j$.
An enhancement of the gravitational collision efficiency due to the turbulent velocity fluctuations can be 
defined as: $E_v=<\frac{|v_1-v_2|}{|w_1-w_2|}>_b$, where $v_i$/$v_j$ is the velocity 
of colliding parcels, either vertical 
only or vertical and horizontal for a $REF2$, and $<>_b$ - averaging operator.  
Although it is possible to calculate this enhancement for each pair of droplets it is easier
to use a bin structure the same as was used to map collisions between parcels, 
and in such a case 
$w_i$/$w_j$ represent a terminal velocity for the center of the bin rather than 
for an individual parcel, and averaging is done for each bin.
Figure \ref{fig9} shows a enhancement of the gravitational collision efficiency.
Four panels show $E_v$ 
for a different bin sizes: 2.6 ${\mu}m$ (microphysical bin 4) - figure \ref{fig9}a, 
9.1 ${\mu}m$ (microphysical bin 10) - figure \ref{fig9}b, 
28.8${\mu}m$ (microphysical bin 16) - figure \ref{fig9}c, 
94.3 ${\mu}m$ (microphysical bin 22) - figure \ref{fig9}d. The largest enhancement is for  small 
bins with a value reaching 125 for a $TURB$ simulation. Simulation $REF2$ has value 120 and REF 
almost 105. Expectedly, the $BIN$ simulation has a constant value of $E_v$ - 1 independent on size, 
because the vertical velocity 
in this case is a droplet terminal velocity. With the increasing droplets sizes the departures 
of enhancement factor from 1 is smaller, and for bin 22 figure \ref{fig10}d it approaches 2. For a given bin, 
the largest $E_v$ is found for adjacent bins. For bins separated by a large distance $E_v$ is 
much smaller. The distribution of the $E_v$ is non-symmetric, 
with larger values of $E_v$ found for the 
sizes smaller than bin under consideration. The results presented in figure \ref{fig9} show that 
even small differences in air velocity can affect velocity statistics for droplets having similar 
sizes and as a result moving with a similar velocities.  For droplets having large differences in 
sizes turbulent velocity would have to be significantly larger than terminal velocity of larger 
droplets to influence $E_v$, which is not a case in the simulations discussed in this article.
The large values of the enhancement of the 
gravitational collision efficiency for small droplets are assiciated with the fact that those droplets adjust 
quickly to the environmental flow, and as a result the velocity of small droplets is very similar to the 
velocity of the air. Much smaller values of the  enhancement of the 
gravitational collision efficiency for large droplets are because these droplets need more time to adjust 
to the environmental flow and their terminal velocity is of the same order as air velocity fluctuations. 
 
Diagnostics show that the maximum turbulent velocity for a $TURB$ case is around 3 m/s and 
has the same order as a terminal velocity of the 650 ${\mu}m$ droplet - $\sim$2.5 m/s. 
Note, however, that 3 m/s represents the largest value and standard deviation of the turbulent 
velocity for the whole cloud is $\sim$0.1-0.25 m/s.
An example of the velocity fluctuations for run $REF$ and $TURB$ are shown in figure \ref{fig10}. 
There is a significant 
difference between the statistics for these two cases. For a $TURB$ case distribution 
of $u^{'}$ and $w^{'}$ are almost identical. For the $REF$ case anisotropy of the 
velocity fluctuations is observed, and tails for a $w^{'}$ distribution
is much broader than for a $u^{'}$, with the statistics for $TURB$ case laying in-between
$u^{'}$ and $w^{'}$ for a $REF$.

\section{Conclusions}
In this article two possible representation of the air turbulent velocity  
in a numerical model with a 
Lagrangian representation of microphysics are discussed.  
Air turbulent velocity in the model is 
represented either as a random walk process, with the standard deviation of velocity 
fluctuations  derived from the diffusion coefficient predicted by the numerical 
model (Eulerian part) or as an interpolation of an air velocity to a parcel location. 
The random walk model is derived for an anelastic approximation 
and it's shown that an additional, deterministic,  term $1/{\rho} \nabla (\rho K_m)$ 
needs to be included in a random walk model for consistency with the Eulerian model. 
It is argued that the mixing length used in a Lagrangian model 
should be smaller than used to derived diffusion coefficient for an Eulerian model;
and the mixing length scale based on TKE and model time-step for use in Lagrangian 
microphysics is introduced. It is demonstrated that interpolation of the velocity
to parcel location and use of this velocity in a collision kernel has a similar effect 
to representation of the turbulence as a random walk, and can be treated as an 
alternative to a random walk model. 
Additionally, for turbulence treated as a random 
walk, because all parcels within the particular computational grid have the same value 
of the deterministic velocity, fluctuations in the parcel number in a 
computational grid may be larger than for the case when velocity is interpolated 
to a parcel location. Unlike the random walk model,
interpolation takes into account possible anisotropy
in the flow velocity, also observed in a laboratory studies 
\citeauthor{malinowskiNJP}~(\citeyear{malinowskiNJP}), \citeauthor{korczyk}~(\citeyear{korczyk}) but in a much smaller scale.

If the sub-grid scale transport and droplet collisions occur on the same scale,  air turbulent
velocity can significantly enhance velocity differences between droplets, especially those
having small sizes, when the turbulent velocity is much larger than terminal velocity of 
these droplets. In the cases discussed in this article enhancement as large as 120 has 
been observed for the case when only vertical velocity is taken into account when 
calculating relative velocity of colliding parcels. Allowing for the differences
in a horizontal velocity
$E_v$ increases by $\sim$15 \%.
The turbulent enhancement of the velocity of the colliding droplets obtained in this article
is much larger than obtained form a DNS simulations and used in LES 
model by \citeauthor{andrzej}~(\citeyear{andrzej}) or \citeauthor{khain2012}~(\citeyear{khain2012}). This difference is associated with the fact that in
Lagrangian microphysics the same velocity is used for the transport 
and when evaluating collision kernel. The values obtained with air turbulent velocity 
fluctuations ($REF$, $TURB$) provide an upper limit of the impact of the air velocity fluctuations on gravitational
collision efficiency, with the lower limit given by values from the BIN simulation. An additional 
parameterization is needed to accont for scale separation between transport and
droplet interactions.

\begin{figure*}[tbh]
\centerline{\includegraphics[width=6.0in,clip=true]{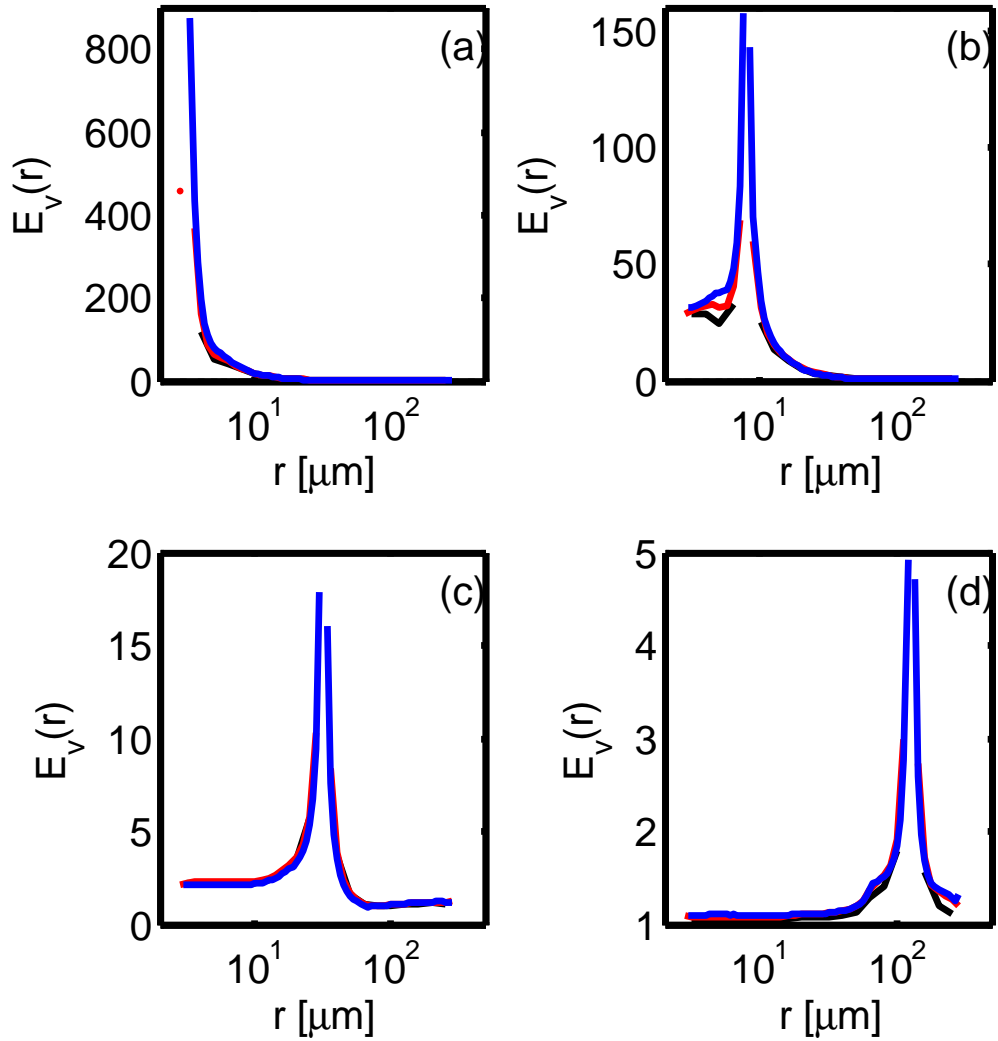}}
\caption{A velocity enhancement factor (E$_v$) for a different microphysical grids:
for radius  2.6 ${\mu}m$  panel a,
9.1 ${\mu}m$ - panel b,
28.8${\mu}m$ - panel c,
94.3 ${\mu}m$ - panel d, for simulation:
$REF2$ - black, $REF2$ with 55 bins used to represent droplet domain - red, 
$REF2$ with 109 bins used to represent droplet domain - blue.
}
\label{fig11}
\end{figure*}

Velocity enhancement diagnosed from the model is non-symmetric, with larger values for
sizes smaller than size under consideration. The asymmetry, however, may be related
to the microphysical grid, because in E$_v$ averaged within a bin 
parcel velocity difference is normalized by
a difference in the terminal velocity for bins these parcels belong to. 
As a result of the averaging and normalization E$_v$ also depends on the number 
of bins used to represent 
domain in a radius space. Figure \ref{fig11} shows
E$_v$ for a $REF$ case together with an additional 2 simulations using the same setup, 
but one with 55 bins (p=2$^1/2$),
and other with 108 bins (p=2$^1/4$). With the increasing number of bins E$_v$ 
also increases, reaching the value of 900 when the smallest bin is under consideration. 
A E$_v$ values approach 1 independent on resolution 
in a radius space when the differences in droplet sizes is large.

Turbulent velocity enhancement can affect significantly drizzle formation. 
Simulation $BIN$, where droplet vertical velocity has been set to a terminal velocity,
based on the bin to which droplets belong
produces negligible amount of rain water, 
and much smaller than for other simulations droplet sizes. Note, however, that even 
in this case droplets do collide forming larger ones. 

For simulations including a representation of air turbulen velocity fluctuations,
drizzle forms initially preferentially near the cloud edge,
near the entrainment eddies or near the cloud top, and often in the areas where
q$_c$ is elevated. 
The edge of the cloud is a place
where due to entrainment and mixing droplet spectrum is broader than in the center
of the cloud and as a result coalescence between droplets is more efficient. Additionally, 
because of the gradient of the velocity near the cloud edge, the differences in the
relative velocity of parcels there is large and this also enhances 
probability of droplet coalescence. Formation of the first drizzle near the cloud edge
it a bin model with the representation of the effect of turbulence on droplet collision rate 
has been also reported recently 
by \citeauthor{khain2012}~(\citeyear{khain2012}).

The cloud droplet spectrum is relatively broad (of the order of $\sim1{\mu}m)$ from the onset of 
the cloud formation. This value is much larger than reported for parcel models in the 
past and large enough to trigger the coalescence process even for the case with a very high 
cloud droplet concentration. It's acknowledged, however, that when the standard 
deviation for each computational grid is considered values smaller than 0.1 ${\mu}$m 
are observed, indicating that
very narrow (comparable to a parcel model) droplet distributions also form within  
computational grids.

Coalescence does process aerosol, but the time scale of this process is longer 
than $\sim$20 minutes of cloud evolution discussed here. Processing by multiple clouds
is needed for the boundary layer aerosol
to change the shape of the aerosol distribution.  

\section{Appendix}
Consider conservation equation for a scalar for an anelastic approximation:
\begin{equation}
\frac{\partial\rho C}{\partial t}=-\nabla(\rho vC)+\nabla(\rho K\nabla C),
\end{equation}
where $C$ - is concentration, $\rho$ - gas density, $v$ - gas velocity, 
$K$ - diffusion coefficient (note that in general $K$ is a tensor, however, here 
it is assumed that this tensor has only diagonal values not equal to 0; 
K$_{i,j}$=${\delta}_{i,j}K$). This equation can be written as:
\begin{equation}
	\frac{\partial\rho C}{\partial t}=-\nabla[\rho C (v+\frac{1}{\rho}\nabla(\rho K)]+\triangle(\rho K C).
\end{equation}
This equation corresponds to the following Stochastic Differential Equation in Ito sense 
(see for instance: \citeauthor{Salamon2006}~(\citeyear{Salamon2006}) or \citeauthor{Spivakovskaya}~(\citeyear{Spivakovskaya}) for the case when $\rho$ is constant):
\begin{equation}
dX=(v+\frac{1}{\rho}\nabla(\rho K))dt+\sqrt{2K}dW(t),
\end{equation}
where $dW(t)$ is a Wiener process. Integration of this equation gives:
\begin{equation}
\triangle X=\int_{0}^{\triangle t}(v+\frac{1}{\rho}\nabla(\rho K))dt+\int_{0}^{\triangle t}\sqrt{2K}dW(t),
\end{equation}
or in numerical representation for i-th direction, 
assuming that $v$, $K$ and $\rho$ are constant during the time step:
\begin{equation}
\triangle X_{i}=(v_{i}(X_{i})+\frac{1}{\rho(X_{i})}\frac{\partial}{\partial x_{i}}(\rho(X_{i})K(X_{i}))\triangle t+\sqrt{2K\triangle t}G,
\end{equation}
where $G$ is a random number having normal distribution. 
It follows that effectively turbulent diffusion corresponds to a random walk process 
with mean velocity having two components, first being the velocity from the 
numerical model, the other accounting for a change in the density 
and diffusivity in a non-homogeneous medium:
\begin{equation}
v_{e}=v_{i}(X_{i})+\frac{1}{\rho(X_{i})}\frac{\partial}{\partial x_{i}}(\rho(X_{i})K(X_{i}));
\end{equation}
and a random component:
\begin{equation}
v_{r}=\sqrt{\frac{2K}{\triangle t}}.
\end{equation}
 
To find $K$ in numerical model an assumption must be made about the mixing length $l$,
$K \sim l \cdot TKE^{1/2}$.
Typically it is assumed to be of the order of a grid length. Although in the 
Eulerian model this assumption is justified, because mixing within each model 
grid must be completed within the time-step, in a Lagrangian transport model, 
where the exact location of each parcel is known it's not necessary true. For a 
Lagrangian model a different length scale can be used, for instance from model 
time-step and turbulent kinetic energy (TKE), which is a prognostic variable in a 
sub-grid scale model. The following length scale can be defined: $l_L$=${\triangle t} \sqrt{2TKE/3}$, 
which for a 2D case discussed in this article is equal to $l_L$=${\triangle t} \sqrt{TKE}$. 
It follows that 
\begin{equation}
v_r \sim \sqrt{TKE}.
\end{equation}
This length scale is used only in a Lagrangian model to calculate diffusion coefficient and next
the standard 
deviation of the velocity fluctuations and mean deterministic velocity. 

Diagnostic from the model indicate that the velocity associated with the
variability in $K$ and $\rho$ is very small $\sim$1 mm/s, but this term is
nevertheless kept in the model.

{\footnotesize \setstretch{.5}

}


\end{document}